\documentclass[12pt,preprint]{aastex}

\shorttitle{Imaging Solar Tachocline}
\shortauthors{Zhao et al.}

\begin{document}
\title{Imaging the Solar Tachocline by Time-Distance Helioseismology}

\author{Junwei Zhao\altaffilmark{1}, Thomas Hartlep\altaffilmark{2*},
A. G. Kosovichev\altaffilmark{1}, and N. N. Mansour\altaffilmark{2}}
\altaffiltext{1}{W.W.Hansen Experimental Physics Laboratory, Stanford
University, Stanford, CA~94305-4085}
\altaffiltext{2}{NASA Ames Research Center, Mailstop 230-2, Moffet
Field, CA~94035}
\altaffiltext{*}{present address: W.W.Hansen Experimental Physics 
Laboratory, Stanford University, Stanford, CA~94305-4085}

\begin{abstract}
The solar tachocline at the bottom of the convection zone is an important
region for the dynamics of the Sun and the solar dynamo. In this
region, the sound speed inferred by global helioseismology exhibits a bump
of approximately $0.4\%$ relative to the standard solar model. Global
helioseismology does not provide any information on possible latitudinal
variations or asymmetries between the Northern and Southern hemisphere.
Here, we develop a time-distance helioseismology technique, including
surface- and deep-focusing measurement schemes and a combination of both, 
for two-dimensional tomographic imaging of the solar tachocline that infers
radial and latitudinal variations in the sound speed. 
We test the technique using artificial solar oscillation 
data obtained from numerical simulations. The technique successfully 
recovers major features of the simplified tachocline models. The technique 
is then applied to SOHO/MDI medium-$\ell$ data and provides for the
first time a full two-dimensional sound-speed perturbation image of the
solar tachocline. The one-dimensional radial profile obtained by 
latitudinal averaging of the image is in good agreement with the previous 
global helioseismology result. It is found that the amplitude of 
the sound-speed perturbation at the tachocline varies with latitude, 
but it is not clear whether this is in part or fully 
an effect of instrumental distortion. Our initial results demonstrate 
that time-distance helioseismology can be used to probe the deep
interior structure of the Sun, including the solar tachocline.
\end{abstract}

\keywords{Sun: helioseismology --- Sun: Activity --- Sun: Interior
--- Sun: Magnetic Fields }

\section{Introduction}
The solar tachocline, first defined by \citet{spi92}, is a region located
between the solar convection zone and the radiative zone. It is a transition
zone where the latitudinal differential rotation of the convection
zone changes to a rigid rotation of the radiative zone, as found
by numerous helioseismology studies
\citep[e.g.,][]{tho96,sch98}. In this region the 
inferred sound-speed perturbation profile exhibits the largest difference
relative to a standard solar model - Model S \citep{chr96}, as revealed
by a number of helioseismology studies \citep[e.g.,][]{gou96,
kos97}. The solar tachocline is also widely believed to be the location
where the solar dynamo operates, since the sharp rotational gradient
observed in this area can build up a strong toroidal field
\citep[e.g.,][]{par93, cha97, dik99}. Therefore, it is important
to measure the structure, dynamics, and cyclic variations
of the tachocline. Measurements in this zone will provide a better 
understanding of the transition 
from the solar convection zone to the radiative zone, the mechanisms of
solar magnetic field generation and amplification, as well as
the periodicity and evolution of the solar activity cycles.

However, despite the importance of the tachocline, little is known
about this area from observations besides the steep
rotational shear and the one-dimensional sound-speed perturbation
bump. This region is difficult to probe by helioseismology because 
it is quite thin, $\sim 0.05 - 0.09 R_\sun$  \citep{kos96, bas97}, and 
is located deep in the Sun, approximately $0.29 R_\sun$ below the
photosphere \citep{chr91}. Nearly all observational results related
to the tachocline area were obtained by global helioseismology
analysis of normal mode frequencies, which is not capable of 
resolving hemispheric asymmetries of the differential rotation and  
sound-speed structures. In this paper, we explore the possibility of 
mapping a two-dimensional image of the tachocline, both latitude
and depth dependent, by the use of time-distance helioseismology
\citep{duv93}, a local helioseismology technique.

Because helioseismology infers solar interior properties that
are undetectable by other independent observations, it is often
difficult to assess the quality and accuracy of the inferred properties.
Recently, \citet{geo07} and \citet{zha07a} evaluated the performance
of the time-distance technique for inferring small-scale subsurface
flow fields by analyzing data from numerically simulated solar
convection \citep{ben06}, and found satisfactory results.  
More recently, \citet{har08} validated the time-distance far-side 
imaging technique \citep{zha07b} by using numerical simulations of  
the global acoustic wavefields of the whole Sun. In this paper, we use 
a similar approach: develop time-distance measurement and inversion 
procedures for imaging sound-speed variations in the tachocline,
test these procedures on numerical simulation data, and finally apply the
procedures to actual observations.

To image the solar tachocline, we develop two time-distance measurement
procedures based on surface- and deep-focusing schemes \citep{duv03}, 
and corresponding inversion codes. We also invert combined measurements 
of these two schemes to improve the resolution. A description of 
the measurement and inversion 
techniques is given in Sec.~2. In Sec.~3 we describe the application of 
our method to numerical simulation data, and assess the performance 
of our method. In Sec.~4, we present initial results of the tomographic 
imaging of the solar tachocline using {\sl SOHO}/MDI data \citep {sch95}. 
A brief discussion follows in Sec.~5.

\section{Techniques}
\subsection{Surface- and Deep-Focusing Measurements}
\begin{figure}
\epsscale{0.7}
\plotone{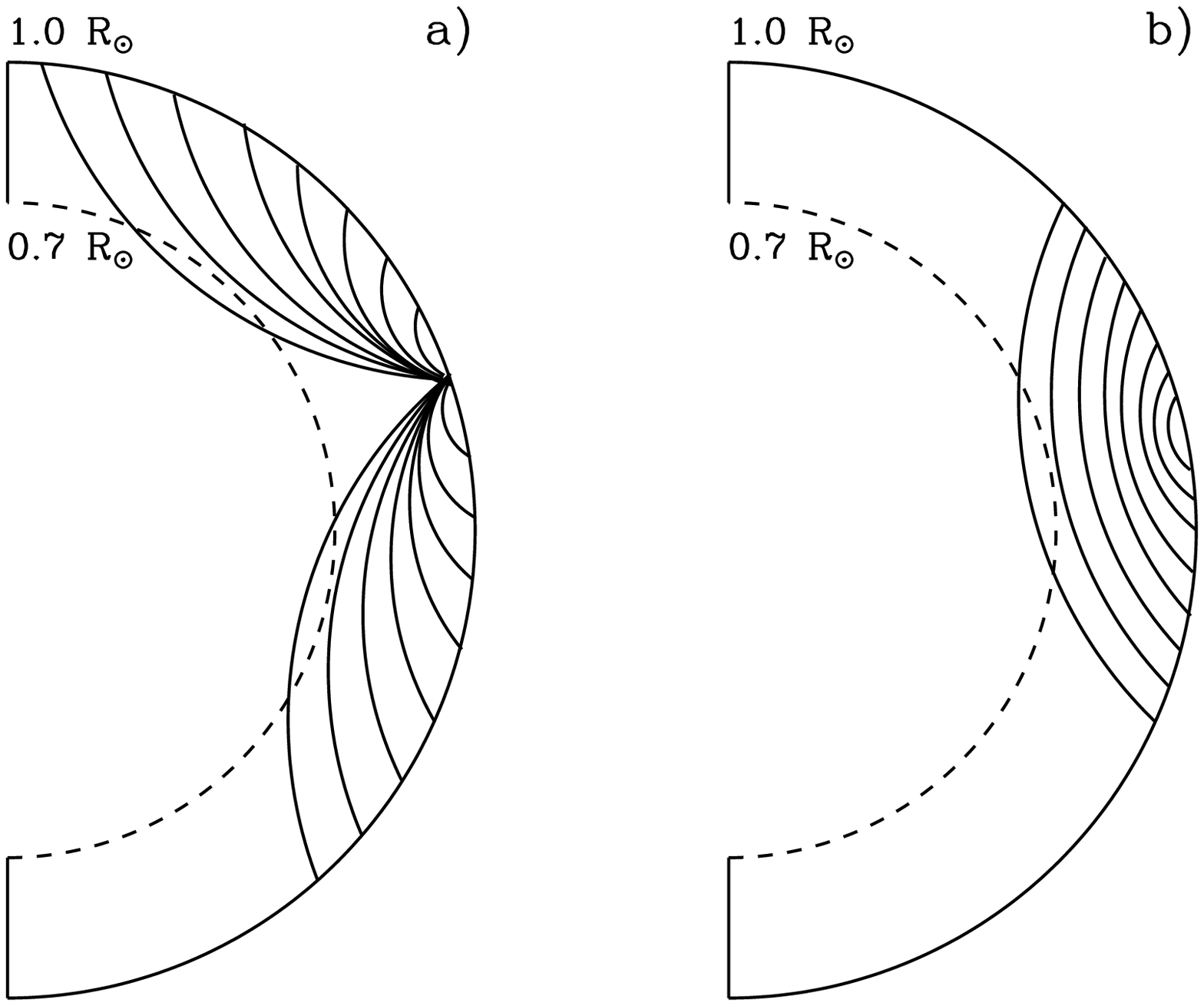}
\caption{Sketches of wave paths in a meridional plane for two measurement 
schemes: a) surface-focusing and b) deep-focusing.}
\label{schemes}
\end{figure}

Time-distance helioseismology measures travel times of acoustic
waves that propagate from one solar surface location to another
along curved paths through the solar interior. These measurements 
are done by computing the cross-correlation between acoustic signals 
observed at these two separate locations. In practice, in order to 
improve the signal to noise ratio, the cross-correlations are
calculated between the acoustic signal observed at a given point
and the signal averaged over its surrounding annulus or part of 
the annulus. When all ray paths converge at one surface location
in such measurements, as shown in Fig.~\ref{schemes}a, this measurement 
scheme is often called surface-focusing. When all ray paths pass 
through one point located in the solar interior, this scheme is 
often referred to as deep-focusing (Fig.~\ref{schemes}b).

The measurement procedure is as follows: A datacube of a
time sequence of surface velocity images, either numerically simulated 
or observed, is remapped into heliographic coordinates
(using Postel's projection) with the center of the projected map 
located at the solar disk center and a pixel size of $0\fdg6$ (hereafter, 
degree means heliographic degree), consistent with the MDI
medium-$\ell$ data sampling resolution.
For the surface-focusing scheme every pixel in the dataset serves 
as a central point, and for each central point the acoustic signals 
are averaged over a set of annuli around this point. Each annulus 
corresponds to a particular range of travel distances of acoustic waves. 
The radius of the annuli ranges from $6\fdg0$ to $84\fdg0$
with a step of $0\fdg6$, hence a total of 131 annuli (travel 
distances) are used. This number is significantly greater than 
the number of annuli used in most previous time-distance helioseismology 
measurements. At each location, the cross-correlations are computed 
between the signals at the central point and the signals from each annulus.
To further improve the signal-to-noise ratio, we average
the cross-correlations in overlapping latitudinal zones $5\fdg0$ wide,  
from $-60\degr$ to $60\degr$ with a step of $0\fdg6$. Then, we use
the Gabor wavelet fitting procedure \citep{kosduv97}  to derive the 
mean acoustic phase travel times for all annuli at each latitude.

For the deep-focusing scheme (Fig.~\ref{schemes}b), every pixel 
also serves as a central point, but the annuli are divided into 
four quadrants in the North, East, South, and West directions. 
The acoustic signals are averaged in each quadrant, and 
cross-correlations are calculated between opposite quadrants. 
Then, these two cross-correlations are averaged together. Note 
that for some locations, not both cross-correlations can be obtained, 
since some quadrant annuli may be located beyond the solar limbs or 
poles. In these measurements, the distance 
between the opposite quadrants varies from $6\fdg0$ to $84\fdg0$ 
with a step of $1\fdg2$; hence, a total of 66 travel distances 
is used. Similar to the surface-focusing
scheme, the cross-correlations are averaged in longitudinal belts, 
and the Gabor wavelet fitting is performed to derive the
acoustic travel time for each annulus and each latitudinal belt.

It is important to note that unlike in most other time-distance
helioseismology studies, the phase-speed filtering is not applied 
in this study, because for the 
large distances used in this study, such filtering may introduce 
systematic shifts that are not fully
understood \citep{duv03}. However, as usual, solar convection and 
$f$-modes are removed by filtering.

\subsection{Inversion Algorithm}
After the acoustic travel times are measured an inversion procedure 
is applied to infer the sound-speed perturbations inside the Sun. 
The travel-time sensitivity kernels in spherical coordinates are 
calculated using the ray-path approximation \citep{dsi95}. 
The sensitivity kernels are computed separately for the two different
measurement schemes. First, three-dimensional kernels are
computed for all different annuli, and the
kernels are then averaged along the longitudinal direction to make 
the kernels depend only on latitude and radius. 

The inversions are performed by solving linearized equations for 
the travel time variations in the sense of least squares \citep{kosduv97}. 
Generally, two different inversion algorithms were used in the past in 
time-distance helioseismology, LSQR algorithm \citep{kos96a, zha01} 
and Multi-Channel Deconvolution, also known as MCD \citep[e.g.,][]{jac99, 
cou05}. In this paper, we employ the MCD technique. 

\section{Results from Numerical Simulation Data}
\subsection{Model and Numerical Simulation}
A numerical simulation code has been developed to simulate the 
three-dimensional acoustic
wavefields of the global Sun in spherical coordinates \citep{har08b}.
It solves the linearized Euler equations describing
wave propagation excited by random sources near the surface. In the 
present study, the simulations include oscillations of
spherical degree, $\ell$, in the range from 0 to 170. This is sufficient
for most global scale studies. Comparing with observations obtained
from the MDI medium-$\ell$ program, this simulation code reproduces 
accurately the solar oscillation power spectrum ($k-\omega$ diagram)
except at high frequencies. 
It also reproduces the acoustic travel times measured by the time-distance
technique \citep[see Fig.~1 in][]{har08}.
Global wavefields generated with this simulation code have been
successfully used for validating the time-distance far-side imaging
technique \citep{har08}.

\begin{figure}
\epsscale{0.7}
\plotone{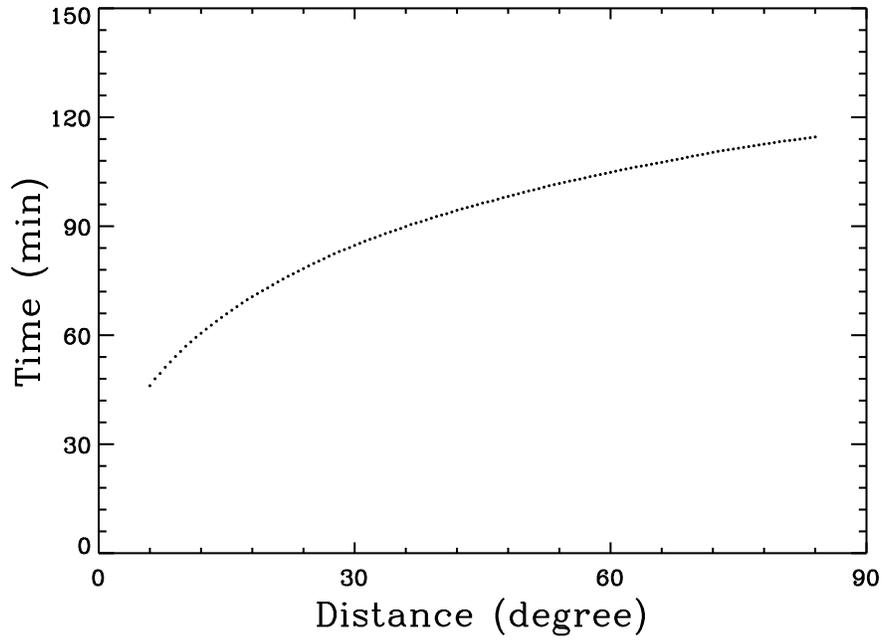}
\caption{Acoustic phase travel time measured from the numerically simulated
global wavefields using solar model~S. The distance step is $0\fdg6$ in
the heliographic coordinate.}
\label{time_dis}
\end{figure}

The sound-speed perturbation near the solar tachocline observed by global 
helioseismology is measured relative to a standard solar model, 
Model~S \citep{chr96}. Therefore, to infer the sound-speed structure at 
the tachocline from the measured acoustic travel times, it is 
necessary to subtract the corresponding reference travel times measured
from a simulation for the unperturbed Model~S. A latitudinally 
averaged reference time-distance
curve obtained for the surface-focusing scheme
is shown in Fig.~\ref{time_dis}. The time-distance relationship
for the deep-focusing scheme is similar.

\begin{figure}
\epsscale{0.8}
\plotone{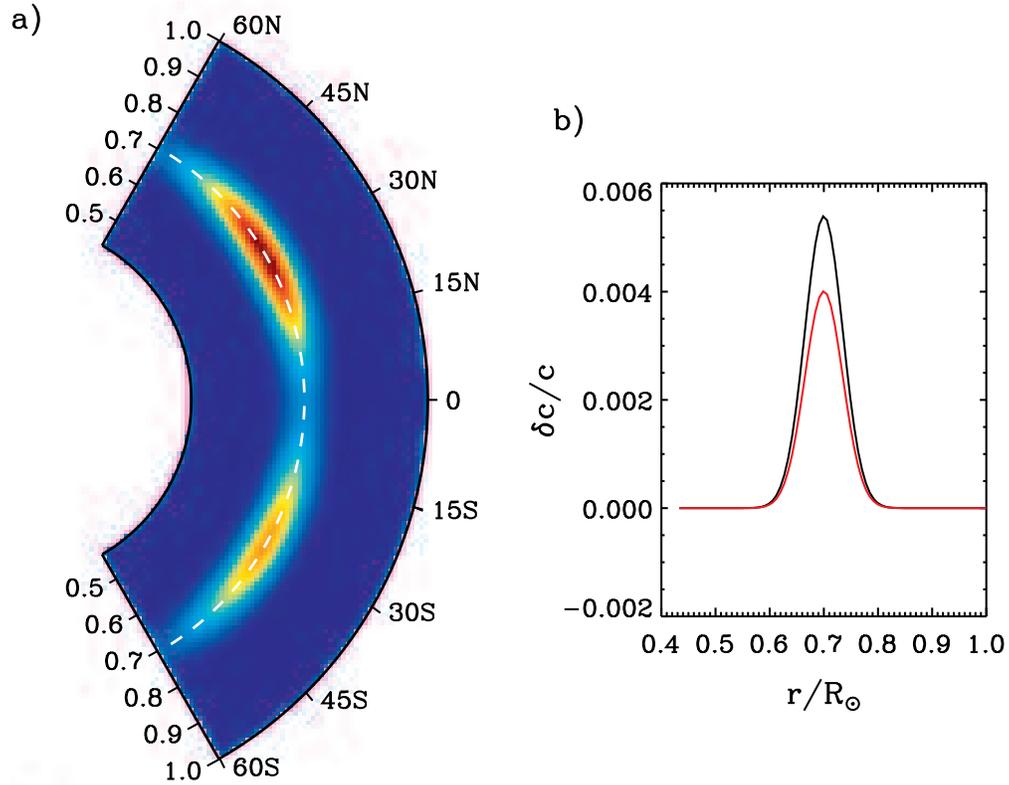}
\caption{a) Sound-speed perturbation model in the tachocline region
used in the numerical simulation of global acoustic wavefields. The 
perturbation maxima, $0.7\%$ and $0.5\%$, are centered at $0.70R_\sun$ and 
latitude of $30\degr$ in Northern and Southern hemispheres, respectively.
The white dashed curve indicates the radius of $0.70R_\sun$ (the bottom of 
the convection zone). b) Latitudinally averaged sound-speed perturbation 
of this tachocline model, with the black curve from the Northern hemisphere, 
and the red from the Southern. }
\label{model}
\end{figure}

For testing, we use an artificial model of the sound-speed perturbations 
in the tachocline shown in Fig.~\ref{model}. The perturbed structure 
has a two-dimensional Gaussian shape in each hemisphere, but is
asymmetric relative to the equator. The maximum amplitudes of sound-speed
perturbation, $0.7\%$ in the Northern hemisphere and $0.5\%$ in the 
Southern hemisphere, are not far apart from the values derived from global
helioseismology \citep[e.g.,][]{kos97}. The perturbation in each 
hemisphere is centered 
at $0.70R_\sun$ with a FWHM of $0.082R_\sun$ (the variance of
the Gaussian is $0.035R_\sun$) in the radial direction, and centered
at latitude of $30\degr$ with a FWHM of $35\degr$ (variance of $15\degr$).
The time series from the numerical simulations of the acoustic wavefields for 
both unperturbed and perturbed models are 1024 minutes long.

\subsection{Surface-Focusing Results}
\begin{figure}
\epsscale{0.7}
\plotone{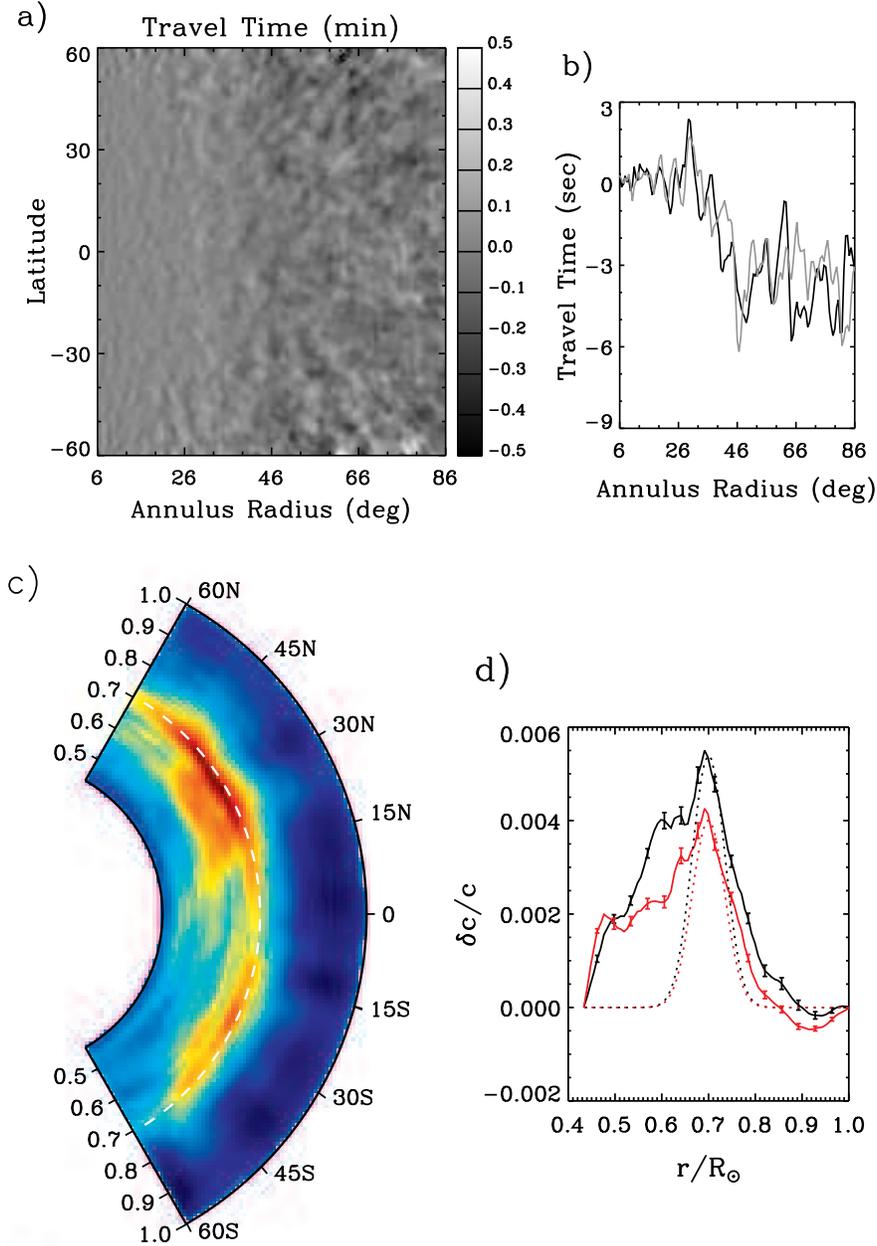}
\caption{Measurements and inversion results for the surface-focusing measurement
scheme: a) acoustic travel times measured from
the numerical simulation with the tachocline model, relative to the 
travel times without tachocline; b) latitudinally averaged relative travel 
times, with dark from the Northern hemisphere and grey from the Southern; 
c) 2D inversion results for the sound-speed perturbation; d) latitudinally 
averaged inversion results, with solid lines showing inversions and dotted 
lines showing the averaged perturbation model. Dark lines are for the 
Northern hemisphere and the red lines are for the Southern.
Errors bars indicate 2$\sigma$ formal error estimates.}
\label{model_surf}
\end{figure}

For the surface-focusing scheme the measured acoustic travel
times after subtraction of the reference travel times are shown 
in Fig.~\ref{model_surf}a as a function of the annulus radius 
(travel distance) and latitude. One can easily see that the travel 
times are apparently smaller on the right-hand
side of the image when the radius is larger than $46\degr$.
This can be better identified in the 1D curves obtained after
latitudinal averaging in both hemispheres, as shown in Fig.~\ref{model_surf}b.

The inversion results for the surface-focusing scheme 
(Fig.~\ref{model_surf}c,d) show that the model sound-speed perturbation 
in the tachocline area is recovered, although it does not perfectly match
the two Gaussian shaped structures in both hemispheres. The location
and amplitude of the perturbation match well. However,
it is obvious from the sound-speed image (Fig.~\ref{model_surf}c) and 
the latitudinally averaged 1D sound-speed perturbation curves
from both hemispheres 
(Fig.~\ref{model_surf}d), that the inverted sound-speed perturbation 
is wider than the prescribed model, especially into 
the deeper interior. A negative sidelobe close to the surface 
can also be seen.

\begin{figure}
\epsscale{0.8}
\plotone{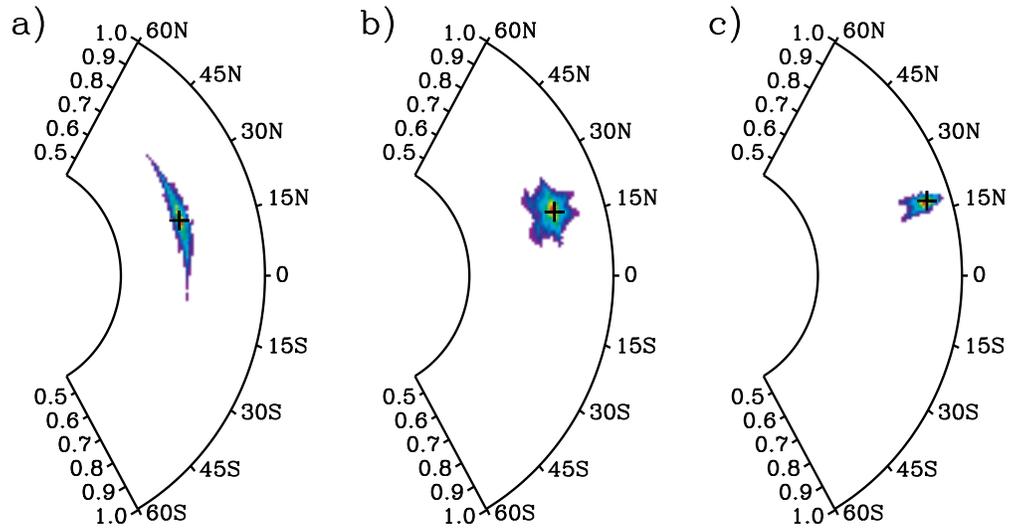}
\caption{A sample of the averaging kernels for the surface-focusing inversions
for the target locations at $18\degr$ latitude and a) $0.70R_\sun$, b) 
$0.80R_\sun$, and c) $0.90R_\sun$. The cross in each panel indicates the 
targeted location. }
\label{kernel_surf}
\end{figure}

It is not immediately clear why the inverted sound-speed results are
not well localized in the deeper area. Inversion averaging kernels, which
characterize the spatial resolution, are displayed in Fig.~\ref{kernel_surf}
for three different targeted depths inside the Sun. It can be seen that when
the target is located near the bottom of the convection zone the averaging
kernel is more spread in latitude, and that the kernel is more
spread in the radial direction when the target location is shallower.
Nevertheless, the averaging kernels are well localized and
should not be the reason for the spreading of the
inverted sound-speed perturbation. We suspect that the poor localization 
with depth below the tachocline may be caused by the use of ray-path sensitivity
kernels and by relatively high realization noise in the simulation data,
which is significantly higher than in the real data since the simulation
time series are shorter. The inversion may be improved
by the use of Born-approximation kernels \citep[e.g.,][]{bir00},
longer time series, and more realistic wave excitation models.

\subsection{Deep-Focusing Results}
\begin{figure}
\epsscale{0.7}
\plotone{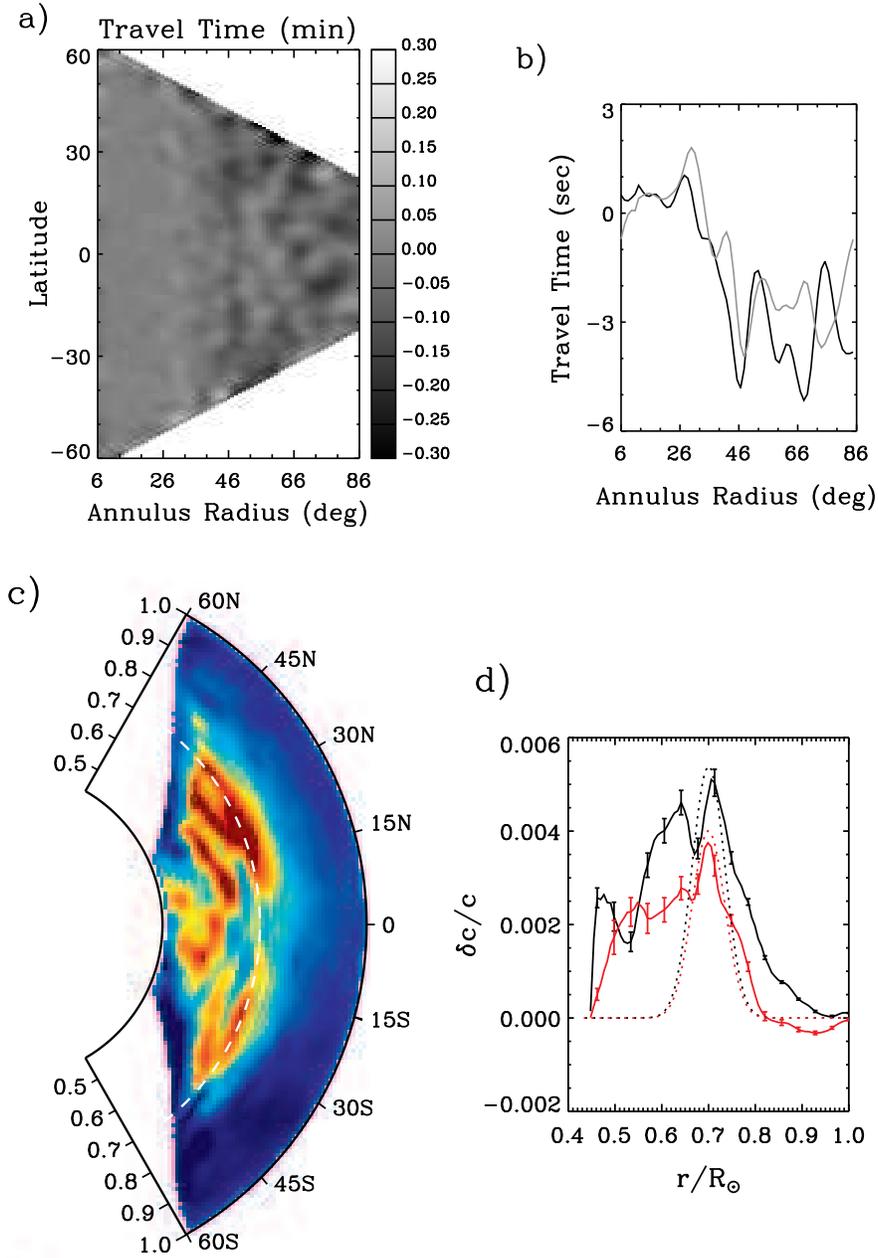}
\caption{Same as Fig.~\ref{model_surf}, but for the deep-focusing
scheme. Note that this scheme provides less coverage than the
surface-focusing scheme.}
\label{model_deep}
\end{figure}

The measurement and inversion results for the deep-focusing scheme are
presented in Fig.~\ref{model_deep}. Similar to the surface-focusing 
measurements, a travel-time drop can be easily identified at an annulus radius
of $\sim 45\degr$ from both the 2D travel-time image (Fig.~\ref{model_deep}a) 
and the 1D latitudinally averaged curves (Fig.~\ref{model_deep}b). 
It is noteworthy that at high latitudes
and large annulus radii, the acoustic travel-time
measurements are often unreliable because too few pixels 
are available for averaging.
The measurements in these areas are discarded to make sure that the 
inversion results are not contaminated by bad fitting points.

\begin{figure}
\epsscale{0.8}
\plotone{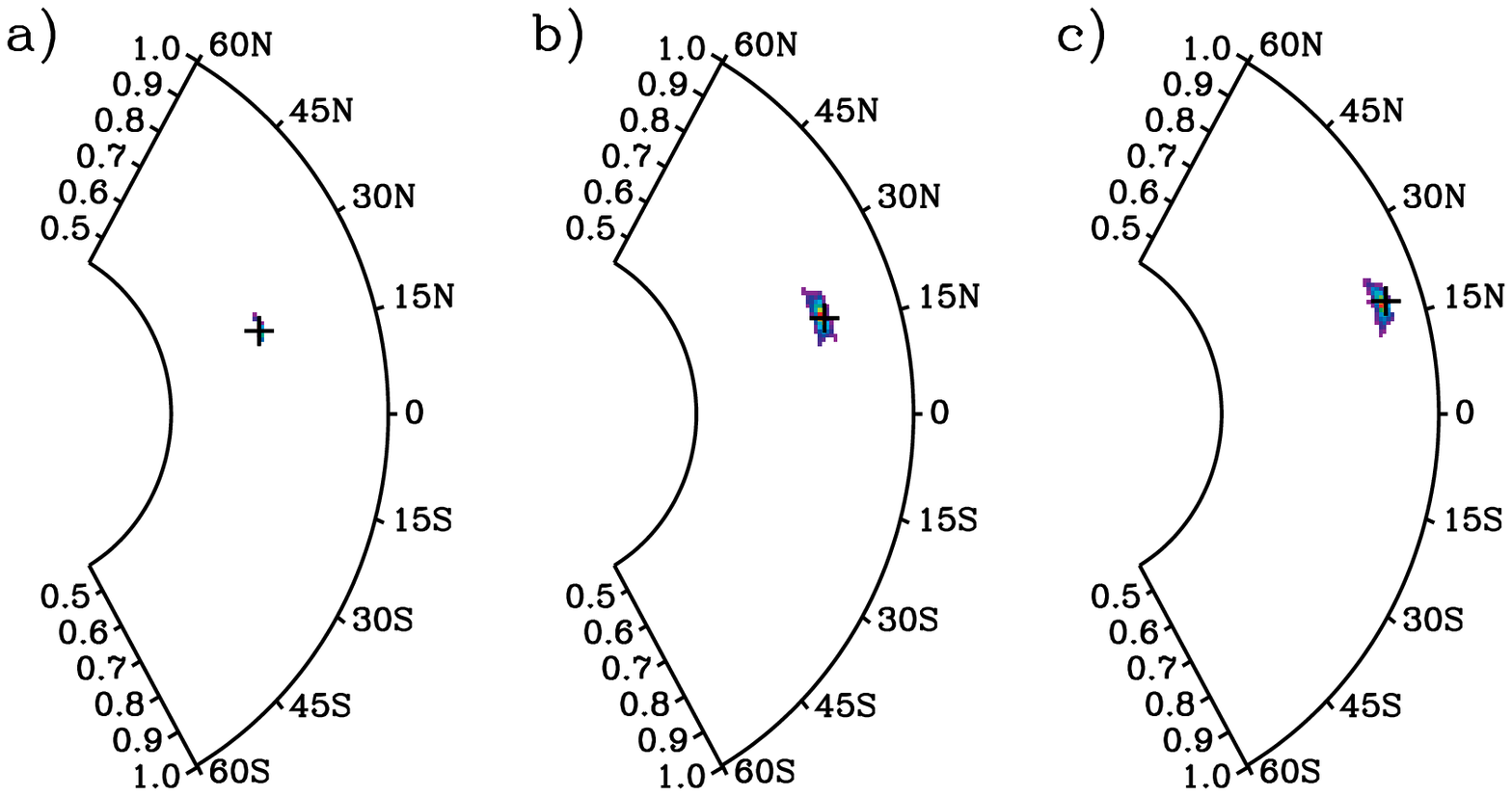}
\caption{Same as Fig.~\ref{kernel_surf}, but for the deep-focusing
scheme.}
\label{kernel_deep}
\end{figure}

The inversion results for the deep-focusing scheme are shown in 
Fig.~\ref{model_deep}c,d. Similar to the surface-focusing inversion
results, the sound-speed bump is successfully recovered, although the shape
does not perfectly match the model. Again, the inferred sound-speed 
perturbation is spread into areas below the tachocline region. 
Despite that the averaging kernels for the deep-focusing inversions 
(Fig.~\ref{kernel_deep}) are well localized, even better than for the 
surface-focusing inversions, the noise level of the inverted results is higher.
Of course, there is a trade-off between localization and noise
amplification in the inversion results. In the model calculations, the trade-off
parameter is chosen to get good localization in the tachocline region.
We did not try to optimize the inversion procedure for the deeper interior. 
The same regularization parameters are used for inversion of the real
data.

\subsection{Results from a Combination of Surface- and Deep-Focusing}
The surface- and deep-focusing measurement schemes use the same 
datasets but are different analysis procedures, and have different 
sensitivities and signal-to-noise ratios at different depths. Thus, 
it is useful to combine the acoustic travel time measurements from 
both schemes, and make one inversion to improve the results. 
It is expected that such joint inversion would improve
localization of the inversion results and enhance the signal-to-noise ratio. 
It is not difficult to revise the inversion code  to solve
two sets of linear integral equations instead of one using a proper combination
of the surface- and deep-focusing sensitivity kernels and measured travel
times. 

\begin{figure}
\epsscale{0.7}
\plotone{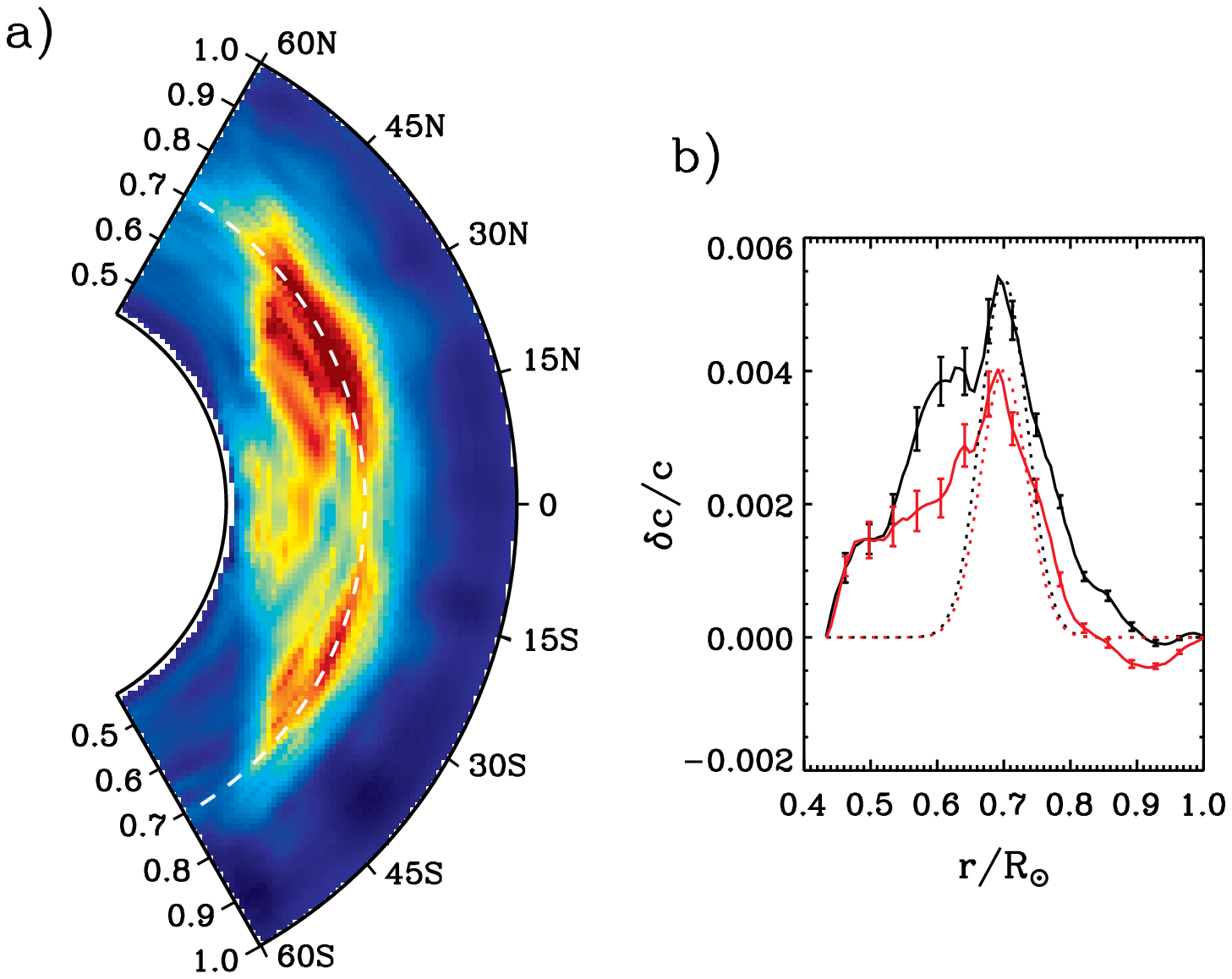}
\caption{Same as panels c) and d) in Fig.~\ref{model_surf}, but for the 
combined inversion of the surface- and deep-focusing measurements.}
\label{model_combine}
\end{figure}

The inversion results are shown in Fig.~\ref{model_combine}. As it can
be seen, both the 2D and 1D results look similar to an average
of the separate inversion results of the surface- and deep-focusing schemes. 
Generally, the inversion from the combined measurements is closer to the model
than each separate inversion. The averaging kernels from this inversion
also look similar to the average of the averaging kernels of these measurement schemes,
and are not shown here.

\section{Results from SOHO/MDI Observations}
\subsection{Data Processing}
The medium-$\ell$ program of the MDI instrument onboard {\sl SOHO} \citep{kos97}
provides nearly continuous space observations of solar oscillation modes 
with angular degree $\ell$ ranging from 0 to
$\sim 300$. The medium-$\ell$ data represent Doppler velocity images
 acquired with 1-minute cadence and a spatial sampling of 
 $10\arcsec$ (0.6 heliographic degrees per
pixel) after some averaging onboard the spacecraft and ground calibration. 
These data can be used for imaging the solar tachocline and
studying its evolution with the solar cycle.

We select the MDI observations from July 25 through August 22, 1996,
covering the whole Carrington rotation CR1912. This period
is during the solar minimum between solar cycles 22 and
 23. This helps to avoid potential complications caused by strong 
surface magnetic fields present during other periods of the
solar cycle. For each day of the observations (1440 minutes long), we remap
the data into heliospheric Carrington coordinates using 
Postel's projection. The main purpose of this remapping 
is to filter out the signals of solar convection and $f$-mode
oscillations. For each observation day, we calculate and save the 
cross-correlation functions for both the surface- and deep-focusing
measurements. Then, the daily cross-correlations are averaged
for the whole Carrington rotation, and Gabor wavelet fitting is performed.
This procedure improves the signal-to-noise ratio. 
The travel-time inversions are done for this Carrington
rotation for the surface- and deep-focusing
schemes separately, and their combination, after the corresponding
reference travel times computed from the unperturbed simulation
(shown in Fig.~\ref{time_dis}) are subtracted.

\subsection{Measurement and Inversion Results}

\begin{figure}
\epsscale{0.7}
\plotone{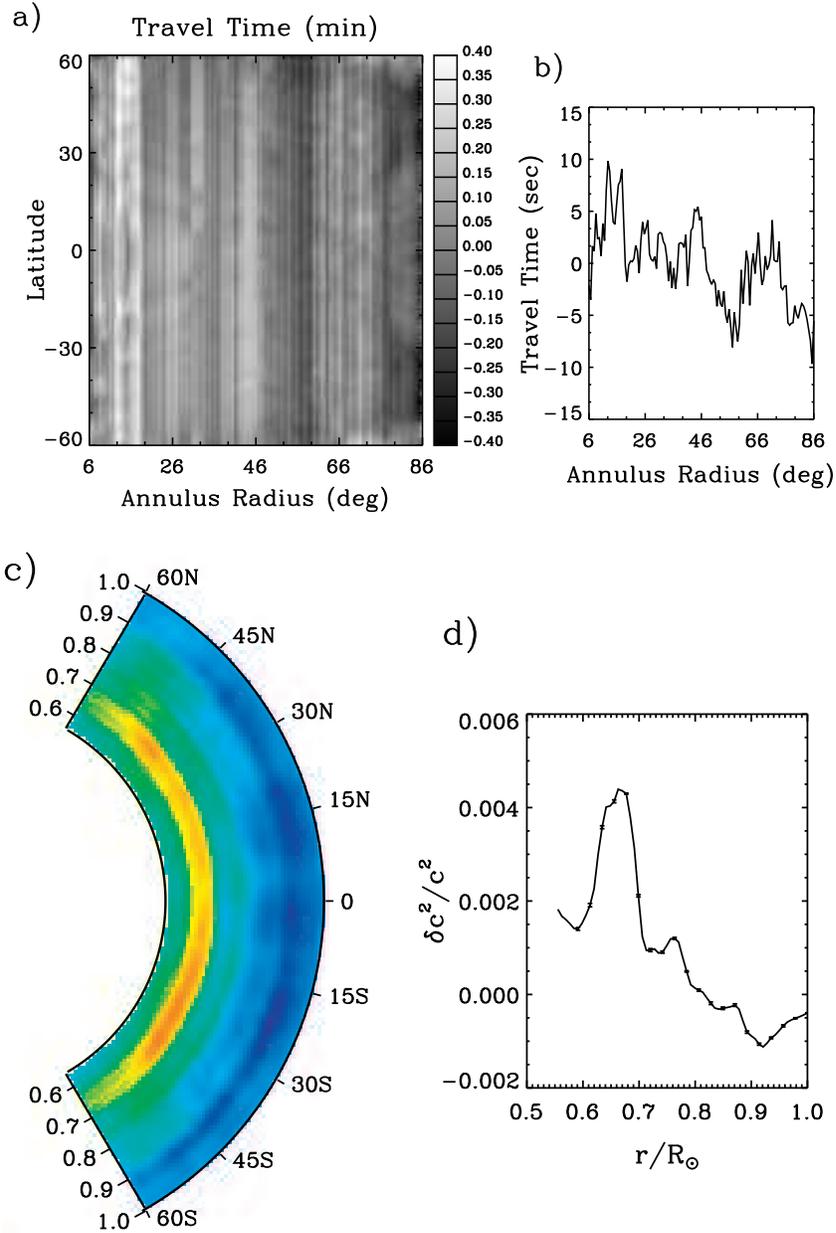}
\caption{Same as Fig.~\ref{model_surf}, but for the SOHO/MDI medium-$\ell$
solar oscillation data and without showing separate hemispheric results.}
\label{real_surf}
\end{figure}

The measured acoustic travel times and inversion results from the
surface-focusing scheme, averaged over the whole Carrington
rotation, are presented in Fig.~\ref{real_surf}. The 2D travel
time image has many thin vertical stripes because
of noise in the reference travel times calculated from
the simulated Sun data. A significant travel time dip can 
be easily identified at the annulus radius between $\sim 45\degr$ 
and $\sim 60\degr$, corresponding to the sound-speed perturbation 
bump near the tachocline region (Fig.~\ref{real_surf}b).
The 2D image of the sound-speed perturbations obtained by inversion 
(Fig.~\ref{real_surf}c) clearly shows a bump at the location of 
approximately $0.67R_\sun$. We find that the perturbation
at the base of the convection zone is not latitudinally uniform, 
being weaker near low latitudes and stronger in latitudes higher than 
about $20\degr$.

\begin{figure}
\epsscale{0.7}
\plotone{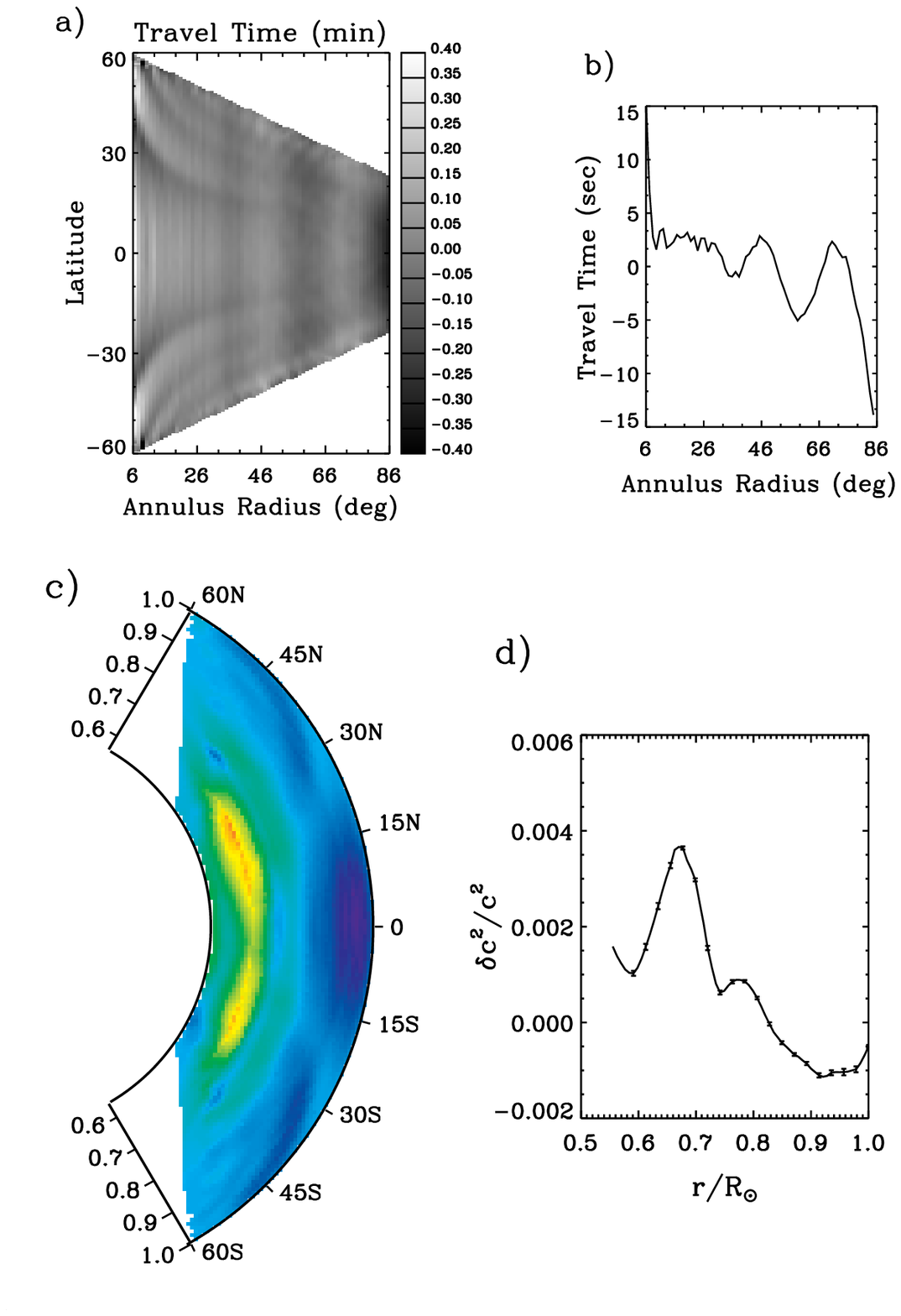}
\caption{Same as Fig.~\ref{real_surf}, but for the deep-focusing scheme.}
\label{real_deep}
\end{figure}

The results for the deep-focusing scheme are presented in Fig.~\ref{real_deep}.
The noise level is significantly larger for this scheme than for 
the surface-focusing scheme. Therefore, we applied a 3-pixel boxcar average 
to the 2D travel-time image (Fig.~\ref{real_deep}a), and used the 
averaged data for inversion. Similar to the surface-focusing results, 
the acoustic travel times
reveal a significant dip for annulus radii from about $46\degr$
to $70\degr$. Besides this dark vertical stripe, one can
also identify in Fig.~\ref{real_deep}a  white and dark curved band 
features in both hemispheres, extending from an annulus radius 
of $6\degr$ to nearly $40\degr$ and between $30\degr$ and $60\degr$ in 
latitude. It is not clear to us whether these features 
are  artifacts or caused by sound-speed perturbations in the Sun. 
The inversion sound-speed results (Fig.~\ref{real_deep}c-d) 
are quite similar to the surface-focusing scheme inferences. 
The sound-speed bump can be clearly identified, and the perturbation 
exhibits an apparent latitudinal dependence as well.

\begin{figure}
\epsscale{0.7}
\plotone{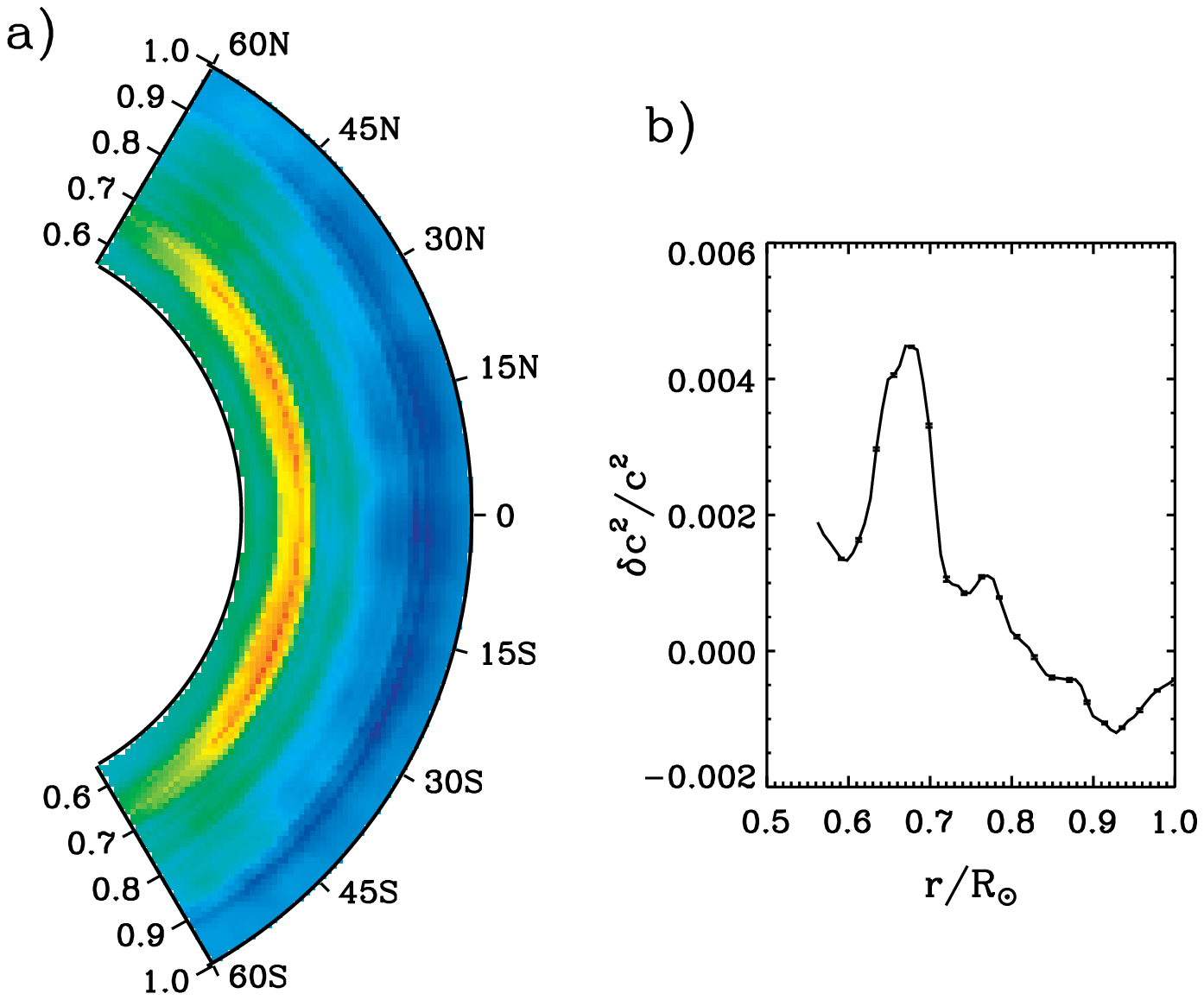}
\caption{Same as panels c) and d) in Fig.~\ref{real_surf}, but for the 
combined inversion of the surface- and deep-focusing measurements.}
\label{real_combine}
\end{figure}

The inversion results from a combination of both surface- and deep-focusing
measurements are shown in Fig.~\ref{real_combine}. Because the
surface-focusing measurements have a better signal-to-noise ratio
and cover higher latitudes than the deep-focusing scheme, we
intentionally give a higher weighting to the surface-focusing measurements
in these inversions. The selection of this weighting parameter 
is quite arbitrary, and the weighting has been chosen as $60\%$ for
the surface-focusing and $40\%$ for the deep-focusing measurement.
It seems that such combined measurements are useful, but may 
still need further optimization.

\begin{figure}
\epsscale{0.8}
\plotone{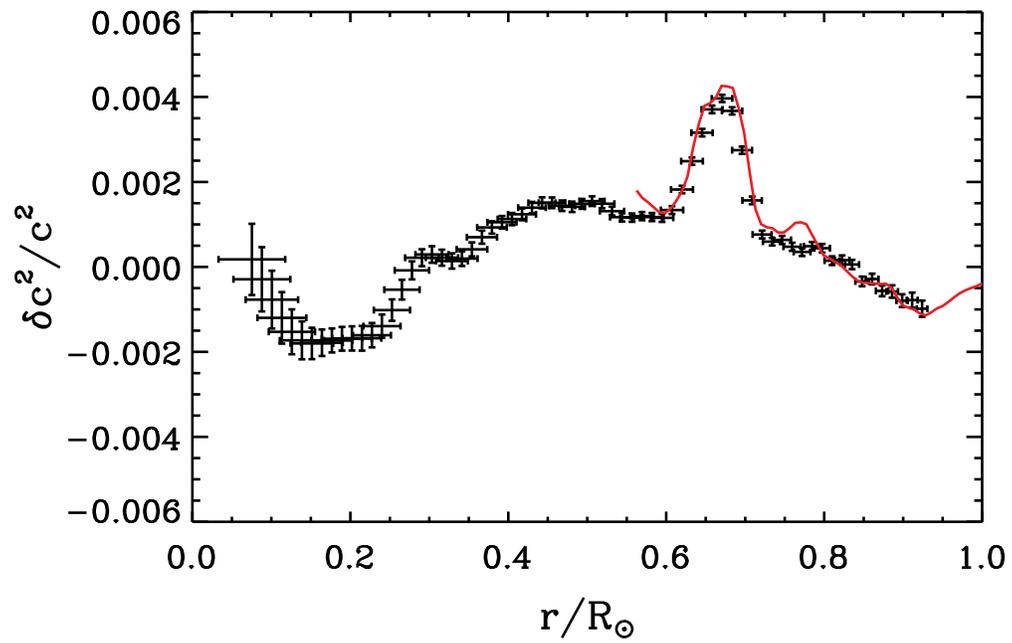}
\caption{Comparison of the sound-speed perturbation obtained from global
helioseismology \citep[symbol with error bars;][]{kos97}, and the
inversion results obtained by combining surface- and deep-focusing
measurements of acoustic travel times (solid curve).}
\label{real_compare}
\end{figure}

It is important to compare our sound-speed profiles with
the results from global helioseismology obtained by inversion
of normal mode frequencies.  Figure~\ref{real_compare} shows 
the comparison of the global helioseismology result obtained 
from the same SOHO/MDI medium-$\ell$ data \citep{kos97}  
with the sound-speed perturbation inferred from the combined
measurements in this study. It is found that both results 
are in good agreement in terms of location and
magnitude of the sound-speed bump near the tachocline, as well
as in the convection zone.

The good agreement of the latitudinally averaged sound-speed
perturbation with the previous global helioseismology result indicates
that our measurements are generally correct. However, the latitudinal
variation of the perturbation is surprising. It certainly requires further
investigation, in particular, of MDI instrumental distortion
\citep{kor04} and non-uniform MTF effects.

\section{Discussion}
Using numerically simulated global acoustic wavefields and solar
oscillation observations from SOHO/MDI, we have demonstrated for
the first time that time-distance
helioseismology is capable of measuring sound-speed perturbations in 
the tachocline located about 200 Mm beneath the solar photosphere
that are relatively small in magnitude and size. This local 
helioseismology method 
provides 2D images of these perturbations as a function of latitude and depth.
This newly developed technique has an advantage over global helioseismology
in that it can distinguish sound-speed perturbations of different amplitude
located at different latitude, and is not limited to the North-South
symmetrical part. Although we only present in this paper 
an example with maximum perturbations of $0.7\%$ and $0.5\%$, our 
experiments demonstrate that the technique is capable of measuring 
the North-South difference in the sound-speed structure of the 
tachocline as small as $0.05\%$.

Our inversions of the numerical simulation data have shown that although
we can recover the major features in the tachocline region, 
the inferred sound-speed perturbations are more widely spread into 
areas beneath the tachocline. We think this may be 
caused by the ray-approximation sensitivity kernels
 used in our inversions or by the relatively high
realization noise in the  simulated wavefield. We plan to
continue the work on improving the resolution of the deep interior
beneath the tachocline.

Our initial analysis of the MDI data shows a latitudinal dependence of 
sound-speed perturbation at the tachocline. We believe this needs further 
investigation and modeling of potential instrumental effects. 
Furthermore, it will be extremely useful to investigate how
the tachocline varies with the solar cycle, and we plan to continue 
these studies in this direction.

\clearpage

\clearpage

\clearpage

\clearpage

\clearpage

\clearpage

\clearpage

\clearpage

\clearpage

\clearpage

\clearpage

\clearpage

\end{document}